\documentclass[11pt]{article}

\usepackage[top=70pt,bottom=80pt,left=100pt,right=100pt]{geometry}

\usepackage{amsmath, amssymb, amsfonts, latexsym, euscript, mathalfa, mathrsfs}
\usepackage{graphicx, color, epsfig, xcolor}
\usepackage{cite}
\usepackage{empheq}
\usepackage{pictexwd,dcpic}

\usepackage{setspace, sectsty}

\usepackage{braket, cancel}

\usepackage[debug,pageanchor=false]{hyperref}
\definecolor{link}{rgb}{.8,.15,.1}
\hypersetup{colorlinks=true,linkcolor=link,citecolor=link,urlcolor=link,linktocpage}

\makeatletter
\@addtoreset{equation}{section}
\makeatother

\newlength{\sswidth}

\newcommand{\nn}{\nonumber}

\def\be{\begin{equation}}
\def\ee{\end{equation}}
\newcommand{\eq}[1]{\begin{equation}\begin{split}#1\end{split}\end{equation}}

\def\e{\epsilon}
\def\g{\gamma}

\def\m{m}
\def\z{\zeta}
\def\o{\omega}
\def\O{\Omega}
\newcommand{\s}{\sigma}
\renewcommand{\t}{\theta}

\newcommand{\vf}{\varphi}

\newcommand{\zbb}{\mathbb{Z}}

\newcommand{\vol}{\mathrm{vol}}

\newcommand{\I}{\text{Im}}
\newcommand{\R}{\text{Re}}

\begin{document}

\begin{titlepage}

\begin{flushright} \small

\end{flushright}

\begin{center}

\noindent

{\Large \bf On supersymmetric AdS$_3$ solutions of Type II}

\bigskip\medskip

Achilleas Passias$^1$ and Dani\"{e}l Prins$^{2,3}$ \\

\bigskip\medskip
{\small

$^1$ Laboratoire de Physique de l’Ecole normale supérieure, ENS, Université PSL, CNRS, Sorbonne Université, Université de Paris, F-75005 Paris, France
\\	
\vspace{.3cm}
$^2$Institut de Physique Th\'{e}orique, Universit\'{e} Paris Saclay, CNRS, CEA, \\
F-91191 Gif-sur-Yvette, France
\\	
\vspace{.3cm}
$^3$Dipartimento di Fisica, Universit\`a di Milano--Bicocca, \\ Piazza della Scienza 3, I-20126 Milano, Italy \\ and \\ INFN, sezione di Milano--Bicocca
}


\vskip .9cm 
     	{\bf Abstract }
\vskip .05in
\end{center}
We classify supersymmetric warped AdS$_3 \times_w M_7$ backgrounds of Type IIA and Type IIB supergravity with non-constant dilaton, generic RR fluxes and magnetic NSNS flux, in terms of a dynamic $SU(3)$-structure on $M_7$. We illustrate our results by recovering several solutions with various amounts of supersymmetry. The dynamic $SU(3)$-structure includes a $G_2$-structure as a limiting case, and we show that in Type IIB this is integrable.
\vfill

\eject

\end{titlepage}

\tableofcontents

\section{Introduction}

Technically more approachable, but not less physically interesting than their higher-dimensional counterparts, AdS$_3$/CFT$_2$ dualities provide a hospitable environment for finding answers to questions on both sides of the holographic correspondence. Conformal field theories in two dimensions, which underlie string theory and are key tools in the description of critical phenomena, feature a highly-constraining infinite-dimensional algebra of conformal transformations that often allows for their exact solution. Gravity in three-dimensional asymptotically anti-de Sitter spacetime provides a toy model for quantum gravity. There is thus a clear motivation for the study of AdS$_3$ backgrounds of string theory.

Owing to the high dimensionality of the internal space, the problem of exploring the space of AdS$_3$ backgrounds is challenging. A way forward is to impose a symmetry on the background, in the form of supersymmetry or isometry, at the expense of the size of the subspace of backgrounds one can access, depending on the degree of the symmetry.
In the present work we impose the minimal amount of supersymmetry, aiming for a more comprehensive scan of supersymmetric AdS$_3$ backgrounds of Type II supergravity. We classify warped AdS$_3 \times_w M_7$ backgrounds with non-constant dilaton, generic RR fluxes and magnetic NSNS flux.
Minimal supersymmetry equips the internal manifold $M_7$ with a dynamic $SU(3)$-structure, due to the existence of two Majorana
spinors on $M_7$. In the limiting case where the spinors are parallel to each other, the dynamic $SU(3)$-structure corresponds to a $G_2$-structure.
We translate the necessary and sufficient conditions for supersymmetry to restrictions on the torsion classes of the $SU(3)$-structure, and obtain expressions for the supergravity fields in terms of the geometric data. We illustrate our results by recovering several AdS$_3$ solutions with various amounts of supersymmetry. In Type IIB supergravity we take a closer look at the $G_2$-structure limiting case, show that it is integrable, and reduce the problem of finding a solution to not only the supersymmetry equations but also the equations of motion, to a single geometric equation. Generically, the dual superconformal field theories in two dimensions preserve $\mathcal{N}=(0,1)$ supersymmetry, and include well-studied families of theories with higher supersymmetry such as those arising from D3-branes wrapped on Riemann surfaces \cite{Benini:2013cda, Benini:2015bwz, Couzens:2017nnr}, whose duals appear in section \ref{IIBsol}.

The rest of this paper is organized as follows. In section \ref{susy}, we present the supersymmetry
equations as a set of equations for a
pair of polyforms on $M_7$. In section \ref{gstructures}, we review $G_2$- and $SU(3)$-structures in seven dimensions, and parameterize the polyforms in terms of the latter. Sections \ref{IIA} and \ref{IIB} contain our results for Type IIA and Type IIB supergravity respectively.

\section{Supersymmetry equations}
\label{susy}

We consider bosonic backgrounds of Type II supergravity whose spacetime is a warped product AdS$_3 \times_w M_7$, where $M_7$ is a seven-dimensional Riemannian manifold.
The ten-dimensional metric reads:
\eq{
ds^2_{10} = e^{2A} ds^2({\rm AdS}_3) + ds^2(M_7) ~,
}
where $A$ is a function on $M_7$. Preserving the symmetries of AdS$_3$,
the NSNS field-strength $H_{10d}$, and the sum of the RR field-strengths $F_{10d}$ in the democratic formulation of Type II supergravity \cite{Bergshoeff:2001pv}, are decomposed as
\begin{equation}
H_{10d} = \varkappa \vol({\rm AdS}_3) + H ~,
\qquad
F_{10d} = e^{3A} \vol({\rm AdS}_3) \wedge \star_7 \lambda(F) + F ~.
\end{equation}
The magnetic fluxes $H$, and
\begin{equation}
\text{Type IIA:} ~ F = \sum_{p=0,2,4,6} F_p~, \qquad
\text{Type IIB:} ~ F = \sum_{p=1,3,5,7} F_p
\end{equation}
are forms on $M_7$. The operator $\lambda$ acts on a $p$-form $F_p$ as $\lambda(F_p) = (-1)^{\lfloor p/2 \rfloor} F_p$. The RR field-strengths satisfy $d_{H_{10d}}F_{10d}=0$, which decomposes as
\begin{equation}\label{eq:dec-Bianchi}
d_H (e^{3A} \star_7 \lambda(F)) + \varkappa F = 0~,
\qquad
d_H F = 0 ~,
\end{equation}
where $d_H \equiv d - H \wedge$. The first set of equations act as equations of motion for $F$, and the second one as Bianchi identities.

We also decompose the ten-dimensional supersymmetry parameters, $\e_1$ and $\e_2$, under
Spin$(1,2) \times$ Spin$(7) \subset$ Spin$(1,9)$:
\eq{
\e_1 = \zeta \otimes \chi_1 \otimes \left(\begin{array}{c} 1 \\ -i \end{array}\right) ~,
\qquad
\e_2 = \zeta \otimes \chi_2 \otimes \left(\begin{array}{c} 1 \\ \pm i \end{array}\right) ~.
}
The upper sign in $\e_2$ corresponds to Type IIA, and the lower sign to Type IIB. $\chi_{1}$ and $\chi_{2}$ are Majorana Spin$(7)$ spinors; $\zeta$ is a
Majorana Spin$(1,2)$ spinor that solves the Killing equation
\begin{equation}
\nabla_\mu \zeta = \frac{1}{2} \m \gamma_\mu \zeta ~,
\end{equation}
where the real constant parameter $m$ is related to the AdS$_3$ radius $L_{{\rm AdS}_3}$ as $L^2_{{\rm AdS}_3} = 1/m^2$.

The Cliff$(1,9)$ gamma matrices are decomposed as
\begin{equation}
\Gamma_\mu = e^A \gamma_\mu^{(3)} \otimes \mathbb{I} \otimes \sigma_3~,\qquad
\Gamma_m = \mathbb{I} \otimes \gamma_m \otimes \sigma_1~,
\end{equation}
where $\gamma_\mu^{(3)}$ are Cliff$(1,2)$ gamma matrices, $\gamma_m$
are Cliff$(7)$ gamma matrices and $\mu, m$ are spacetime indices. We choose $\gamma_\mu^{(3)}$ to be real, and $\gamma_m$ imaginary and antisymmetric. For more details see the appendix of \cite{Passias:2019rga}.

Necessary and sufficient conditions for supersymmetry are generally given in terms of a set of Killing spinor equations. For AdS$_3$ backgrounds, these can be rewritten in terms of  a pair of bispinors $\psi_\pm$ defined by
\eq{\label{polydef}
\chi_1 \otimes \chi_2^t \equiv \psi_+ + i \psi_- ~.
}
Taking into account the Fierz expansion of $\chi_1 \otimes \chi_2^t$, and by mapping anti-symmetric products of gamma matrices to forms, $\psi_+$ and $\psi_-$ can be treated as polyforms on $M_7$, of even and odd degree respectively. The necessary and sufficient conditions for supersymmetry in terms of differential constraints on these polyforms were derived in \cite{Dibitetto:2018ftj} for Type IIA, and in \cite{Passias:2019rga} for Type IIB.

Supersymmetry imposes
\begin{equation}
2 \m c_- = - c_+ \varkappa ~,
\end{equation}
where $c_\pm$ are constants defined by
\begin{equation}
	c_\pm \equiv e^{\mp A}(||\chi_1||^2 \pm ||\chi_2||^2) ~.
\end{equation}
In what follows we will consider backgrounds with zero electric component for $H_{10d}$, $\varkappa = 0$, and thus $||\chi_1||^2 = ||\chi_2||^2$. In Type IIB, $\varkappa = 0$ can be set to zero by applying an $SL(2,\mathbb{R})$ duality transformation.\footnote{We thank N.\ Macpherson for this observation.} In Type IIA, as shown in \cite{Dibitetto:2018ftj}, $\varkappa \neq 0$ leads to zero Romans mass; such AdS$_3$ backgrounds can thus be studied in M-theory, see \cite{Martelli:2003ki, Tsimpis:2005kj, Babalic:2014fua}.
Without loss of generality,  we set $c_+ = 2$, that is
\begin{equation}
||\chi_1||^2 = ||\chi_2||^2 = e^A ~.
\end{equation}
Given the above choices, the system of supersymmetry equations then reads:
\begin{subequations}\label{SUSY}
\begin{align}
d_H(e^{A-\phi} \psi_\mp) &= 0 ~, \label{SUSYa}\\
d_H(e^{2A-\phi} \psi_\pm) \mp 2 \m e^{A-\phi} \psi_\mp &= \frac{1}{8} e^{3A} \star_7 \lambda(F) ~, \label{SUSYb}\\
(\psi_\mp, F)_7 &= \mp \frac \m2 e^{-\phi} \vol_7 ~. \label{SUSYc}
\end{align}
\end{subequations}
Here, an upper sign applies to Type IIA and a lower one to Type IIB; $(\psi_\mp,F)_7 \equiv (\psi_\mp \wedge \lambda(F))_7$, with
$(\cdot)_7$ denoting the projection to the seven-form component.

We can decompose $\chi_2$ in terms of $\chi_1$ as follows:
\eq{\label{chi_rel}
\chi_2 =  \sin \t  \chi_1 - i \cos \t v_m \g^m \chi_1 ~,
}
where without loss of generality, we take $v$ to be  a real one-form of unit norm and restrict $\t \in [0, \pi/2]$.
At the boundary value $\theta = 0$, $\chi_1$ and $\chi_2$ are orthogonal and define a ``strict''  $SU(3)$-structure on $TM_7$.
At the other boundary value $\theta = \pi/2$,  $\chi_1$ and $\chi_2$  are parallel and define a $G_2$-structure.
At intermediate values of $\theta$, the pair $(\chi_1, \chi_2)$ define a ``dynamic'' $SU(3)$-structure on $TM_7$, or alternatively a $G_2 \times G_2$-structure on the generalized tangent bundle $TM_7 \oplus T^*M_7$.

In the next section we review $G_2$- and $SU(3)$-structures in seven dimensions, and parameterize $\psi_\pm$ in terms of the latter.

\section{$G_2$- and $SU(3)$-structures in seven dimensions}\label{gstructures}

We briefly summarize the mathematical formalism for $G_2$- and $SU(3)$-structures on seven-dimensional Riemannian manifolds that we will use in analyzing the supersymmetry equations \eqref{SUSY}.

A $G_2$-structure on a seven-dimensional Riemannian manifold $M_7$ is defined by a nowhere-vanishing, globally defined three-form $\varphi$.
Equivalently, a $G_2$-structure is defined by a nowhere-vanishing, globally defined Majorana spinor. The three-form $\varphi$ is constructed as a bilinear of the Majorana spinor as
\begin{equation}
\varphi_{mnp} = -i \chi^t \gamma_{mnp} \chi ~,
\end{equation}
where $\chi$ is taken to have unit norm. The  three-form $\varphi$ is normalized so that
\begin{equation}
\varphi \wedge \star_7 \varphi = 7 \vol_7 ~.
\end{equation}

In the presence of a $G_2$-structure, the differential forms on $M_7$ can be decomposed into irreducible representations of $G_2$. In particular, this may be applied to the exterior derivative of the three-form $\varphi$ and its Hodge dual $\star_7 \varphi$:
\begin{subequations}\label{g2torsion}
\begin{align}
d\varphi &=  \tau_0 \star_7 \varphi + 3 \tau_1 \wedge \varphi + \star_7 \tau_3 ~, \\
d \star_7 \varphi &= 4 \tau_1 \wedge \star_7 \varphi + \star_7 \tau_2 ~.
\end{align}
\end{subequations}
The $k$-forms $\tau_k$ are the torsion classes of the $G_2$-structure.
$\tau_0$ transforms in the ${\bf 1}$ representation of $G_2$, $\tau_1$ in the ${\bf 7}$, $\tau_2$ in the {\bf 14}, and $\tau_3$ in the ${\bf 27}$.

An $SU(3)$-structure on a seven-dimensional Riemannian manifold $M_7$ is defined by a nowhere-vanishing, globally defined triplet comprising a real one-form $v$, a real two-form
$J$, and a complex decomposable three-form $\Omega$, subject to the following defining relations:\footnote{In terms of local coordinates,
\begin{equation}
X \lrcorner \o_{(k)} \equiv \frac{1}{k-1!} X^n \o_{n m_1 ... m_{k-1}} dx^{m_1} \wedge...\wedge dx^{m_{k-1}} ~.
\end{equation}
}
\begin{align}
v \lrcorner J = v \lrcorner \Omega = 0~, \quad
\Omega \wedge J = 0 ~, \quad
\frac{i}{8}\Omega \wedge \overline\Omega = \frac{1}{3!} J \wedge J \wedge J~.
\end{align}
Equivalently, an $SU(3)$-structure is defined by a pair of non-parallel Majorana spinors; see the appendix of \cite{Passias:2019rga}.
The one-form $v$ foliates $M_7$ with leaves $M_6$. The metric on $M_7$ is then locally decomposed as
\begin{equation}\label{gM7}
ds^2(M_7) = v^2  + ds^2(M_6) ~,
\end{equation}
with an accompanying volume form $\vol_7 \equiv \frac{1}{3!} v \wedge J \wedge J \wedge J$. The exterior derivative can be decomposed as
\begin{equation}
d = v \wedge v \lrcorner d + d_6 ~,
\end{equation}
where $d_6$ is the exterior derivative on $M_6$.
$k$-forms orthogonal to $v$ can be decomposed into primitive $(p, q)$-forms with respect to $J$.

The intrinsic torsion of an $SU(3)$-structure splits into torsion classes, which transform in irreducible representations of $SU(3)$. They parameterize the exterior derivatives of $(v, J, \Omega)$ as:
\begin{subequations}\label{su3torsion}
\begin{align}
dv &= R J + T_1 + \R (\overline{V_1} \lrcorner \O) + v \wedge W_0~,  \\
dJ &= \frac32 \I ( \overline{W_1} \O) + W_3 + W_4 \wedge J
+ v \wedge \left( \frac23 \R E J  + T_2 + \R (\overline{V_2} \lrcorner \O) \right)~,  \\
d \O &= W_1 J \wedge J + W_2 \wedge J + \overline{W_5} \wedge \O
+ v \wedge \left(E \O - 2 V_2 \wedge J + S \right)~.
\end{align}
\end{subequations}
$R$ is a real scalar, $E$ and $W_1$ are complex scalars, $V_1$, $V_2$ and $W_5$ are complex $(1,0)$-forms, $W_0$ and $W_4$ are real one-forms, $T_1$ and $T_2$ are real primitive $(1,1)$-forms, $W_2$ is a complex primitive $(1,1)$-form, $W_3$ is a real primitive $(2,1)+(1,2)$-form,
and $S$ is a complex primitive $(2, 1)$-form. $R$, $E$ and $W_1$ transform in the ${\bf 1}$ representation of $SU(3)$, $V_1$, $V_2$ and $W_5$ in the ${\bf 3}$, $W_0$ and $W_4$ in the $\bf 3 + \overline{\bf 3}$, $T_1$, $T_2$ and $W_2$ in the ${\bf 8}$, $W_3$ in the ${\bf 6} + \overline{\bf 6}$, and $S$ in the ${\bf 6}$.

As detailed in \cite{Passias:2019rga}, the polyforms $\psi_\pm$ are parameterized in terms of the $SU(3)$-structure as
\eq{\label{polyforms}
\psi_+ &=  \frac{1}{8}  e^A \left[  \I(e^{i \t} e^{iJ}) + v\wedge \R(e^{i \t}\O) \right] ~, \\
\psi_- &=  \frac{1}{8}  e^A \left[ v \wedge \R(e^{i \t} e^{iJ}) + \I(e^{i \t}\O) \right] ~,
}
where $\theta$ is the angle appearing in \eqref{chi_rel}.
When $\theta = \pi/2$, the one-form $v$ drops out of the decomposition \eqref{chi_rel} of $\chi_2$, the spinors $(\chi_1, \chi_2)$ become parallel, and thus define merely a $G_2$-structure rather than an $SU(3)$-structure.
Nevertheless, the above decomposition is still valid: it can be shown that for compact $M_7$, existence of a Spin$(7)$ structure implies existence of an $SU(3)$-structure \cite{Friedrich:1995dp}. Hence, we may decompose the three-form $\varphi$ defined by the spinor $\chi_1 = \chi_2$ in terms of an $SU(3)$-structure $(v, J, \O)$, leading to the above result even in this limiting case.
The phase $e^{i \t}$ multiplying $\Omega$ can be ``absorbed'' by a redefinition $e^{i \t} \O \rightarrow \O$, and we will apply this redefinition in the following sections.

\section{Type IIA}\label{IIA}

In this section we analyze the Type IIA supersymmetry equations \eqref{SUSY} (upper sign): by substituting \eqref{polyforms} in \eqref{SUSY} (redefining $e^{i \t} \O \rightarrow \O$), the necessary and sufficient conditions for supersymmetry translate to restrictions on the torsion classes of the $SU(3)$-structure defined by $(\chi_1, \chi_2)$ on $M_7$. Furthermore, the RR and NSNS field-strengths are expressed in terms of the $SU(3)$-structure data.

The geometry of $M_7$ and the NSNS field-strength are constrained by equation \eqref{SUSYa}, which yields:\footnote{$J^2 \equiv J \wedge J$ and $J^3 \equiv J \wedge J \wedge J$.}
\begin{subequations}\label{IIANSNS}
\begin{align}
d\left(e^{2A- \phi}\cos\t v\right) &= 0 ~, \label{IIAdv} \\
d\left(e^{2A- \phi}\left(- \sin\t v \wedge J + \I \O\right)\right) - e^{2A-\phi} \cos\t H \wedge v &= 0 ~, \\
d\left(e^{2A- \phi} \cos\t v \wedge J^2\right) + 2 e^{2A-\phi} H \wedge \left(- \sin\t v \wedge J + \I \O \right) &= 0
~.
\end{align}
\end{subequations}
The RR field-strengths are derived from \eqref{SUSYb}, corresponding to
\begin{subequations}\label{IIARstar}
\begin{align}
e^{3A} \star_7 F_6 &= - d\left(e^{3A-\phi}\sin\t\right) + 2 \m e^{2A - \phi} \cos\t v ~,  \\
e^{3A} \star_7 F_4 &= d\left( e^{3A- \phi} \cos\t J\right) - e^{3A-\phi} \sin\t H - 2 \m e^{2A-\phi} \I \O \nn \\
&+ 2m e^{2A-\phi} \sin\t v \wedge J ~, \\
e^{3A} \star_7 F_2  &= -d\left(e^{3A-\phi}\left(v \wedge \R\O- \sin\t \tfrac{1}{2} J^2 \right)\right) + e^{3A- \phi} \cos\t H \wedge J \nn \\
&- \m e^{2A- \phi} \cos\t v \wedge J^2 ~,  \\
e^{3A} \star_7 F_0 &= - d\left( e^{3A-\phi} \cos\t \tfrac{1}{3!} J^3\right) - e^{3 A - \phi} H \wedge \left(v \wedge \R \O - \sin\t\tfrac{1}{2}J^2\right) \nn \\
&-m e^{2A-\phi} \sin\t v \wedge \tfrac{1}{3} J^3 ~.
\end{align}
\end{subequations}

From \eqref{IIANSNS}, using \eqref{su3torsion},  we obtain the following relations for the torsion classes of the $SU(3)$-structure:
\begin{subequations}\label{IIAtorsion}
\begin{align}
0 &= R = V_1 = T_1 = \I W_1 = \I W_2 = \I E ~, \\
0 &= d_6(2A-\phi)^{(1,0)} + W_5 ~, \\
0 &= \cos\theta d_6(2A-\phi) - \sin\t d_6 \t - \cos\theta W_0 ~.
\end{align}
\end{subequations}
Furthermore, using  the decomposition of the NSNS field-strength $H$ with respect to the
$SU(3)$-structure
\eq{\label{Hdecomp}
H &= H^R \R\O + H^I \I\O + \left( H^{(1,0)} + H^{(0,1)} \right)\wedge J + H^{(2,1)} + H^{(1,2)} \\
&+ v \wedge  \left( H_v^{(1,1)}  + H^0_v J + H^{(0,1)}_v \lrcorner \O + H^{(1,0)}_v \lrcorner \overline{\O} \right)~,
}
where $H^{(2,1)}$ and $H_v^{(1,1)}$ are primitive, we obtain
\begin{subequations}\label{IIAH}
\begin{align}
2\cos\t \R(H^{(2,1)}) &= - \I S - \sin\t W_3 ~, \\
2 \cos\t \R(H^{(1,0)}) &= \sin\t \left[-W_4 + W_0 - d_6(2A-\phi)\right] \nn \\
& + 2 \I V_2  - \cos\t d_6 \theta ~, \\
\cos\t H^I &= - \R E - 2\dot{A} + \dot{\phi} - \frac{3}{2} \sin\t \R W_1 ~, \\
H^R &= 0 ~, \\
2 \sin\t \R(H^{(1,0)}) + 4 \I(H_v^{(1,0)}) &= \cos\t W_4 ~.
\end{align}
\end{subequations}
Here we have introduced the notation $\dot{f} \equiv v \lrcorner df$ for any function $f$.
Using \eqref{IIAtorsion} and \eqref{IIAH}, as well as the identities in the appendix of \cite{Passias:2019rga} to Hodge dualize, we derive the following expressions for the RR field-strengths:
\begin{subequations}\label{IIAR}
\begin{align}
e^\phi F_0 &= -\cos\theta(3 \dot{A} - \dot{\phi} + 2 \R E) + \sin\theta(3H_v^0 + \dot{\theta} - 2m e^{-A}) - 4 H^I ~, \\
e^{\phi} F_2 &= X_2 \lrcorner \I \O - \sin\t T_2 - \cos\t H_v^{(1,1)} - \R W_2
- 2 v \wedge \I (Y_2^{(1,0)}) \nn \\
& + \left[2\R W_1 + \cos\t(2H_v^0 + \dot{\theta} - 2me^{-A}) + \sin\t(3\dot{A} - \dot{\phi} + \tfrac{4}{3} \R E) \right] J ~, \\
e^\phi F_4 &=  v \wedge 2\I(\cos\theta W_3^{(2,1)}-  \sin\t H^{(2,1)}) - v \wedge X_4 \lrcorner J \wedge J \nn \\
& + \left(\tfrac32 \cos\theta \R W_1 - 2 m e^{-A} - \sin\t H^I \right) v \wedge \R \O \nn \\
&+ \left[\cos\t(3\dot A - \dot \phi + \tfrac23 \R E)+ \sin\t(2me^{-A}-\dot\theta-H_v^0)\right] \tfrac{1}{2} J^2 \nn \\
&- (\cos\t T_2 -\sin\t H_v^{(1,1)} )\wedge J + \I \left[(\cos\t \overline{V_2} -  2 \sin\t H_v^{(0,1)}) \wedge \O \right] \\
e^\phi F_6 &= \left[ \cos\theta(2m e^{-A}-\dot\t)  - \sin\theta (3 \dot A - \dot \phi) \right] \tfrac{1}{3!} J^3 + v \wedge J^2 \wedge \I (X_6^{(1,0)}) ~,
\end{align}
\end{subequations}
where
\begin{subequations}
\begin{align}
X_2 &\equiv - d_6(3A-\phi) + W_0 - \overline{W_5} - W_5 ~, \\
Y_2 &\equiv \sin\t \left[d_6(3A-\phi) + 2 W_4\right] + \cos\t d_6\t + 2\cos\t(H^{(1,0)} + H^{(0,1)}) ~,\\
X_4 &\equiv \cos\t(d_6A + W_0 + W_4) - \sin\t (H^{(1,0)} + H^{(0,1)}) ~, \\
X_6 &\equiv \sin\t d_6(3A-\phi) + \cos\t d_6\t ~.
\end{align}
\end{subequations}
Substituting the above expressions in the pairing equation \eqref{SUSYc} yields the additional scalar constraint:
\begin{equation}\label{IIApairing}
3 H_v^0 - 6me^{-A} + 2 \dot\t + 6  \cos\theta \R W_1 - 4 \sin\theta H^I  = 0 ~.
\end{equation}

Equations \eqref{IIAtorsion}, \eqref{IIAH}, \eqref{IIAR}, and \eqref{IIApairing} constitute necessary and sufficient conditions for the
preservation of supersymmetry.

\subsection{Solutions}

We now look at solutions of the supersymmetry conditions we have derived. In particular we will recover the $\mathcal{N} = 8$ supersymmetric AdS$_3 \times S^6$ solution of \cite{Dibitetto:2018ftj} (realizing the $F(4)$ superalgebra), and the $\mathcal{N} = (4,0)$ supersymmetric AdS$_3 \times S^3 \times S^3 \times S^1$ solution of \cite{Macpherson:2018mif}. In addition to the supersymmetry equations, the equations of motion are solved provided that the fluxes satisfy the Bianchi identities (see for example \cite{Prins:2013wza}), and this is the case for the solutions below.

\subsubsection*{AdS$_3 \times S^6$ with $\mathcal{N}=8$ supersymmetry}

The angle $\theta$, the warp factor $A$, and the dilaton $\phi$ satisfy
\begin{equation}\label{tAp}
d_6 \theta = 0 ~, \quad d_6 A = 0 ~, \quad d_6 \phi = 0 ~.
\end{equation}
The one-form $v$ is closed -- see \eqref{IIAdv} given \eqref{tAp} -- and locally a coordinate $z$ can be introduced such that
\begin{equation}
v = \xi(z) dz ~,
\end{equation}
for a function $\xi(z)$ which can be fixed by a change of coordinate. Following \cite{Dibitetto:2018ftj}, it is fixed to
\begin{equation}
\xi(z) = - \frac{2}{3} \left(\frac{q}{p}\right)^{1/3} e^{-A(z)} ~,
\qquad p,\,q = {\rm constants} ~.
\end{equation}
Accordingly, the metric on $M_7$ \eqref{gM7} reads
\begin{equation}
ds^2(M_7) = \frac{4}{9} \left(\frac{q}{p}\right)^{2/3} e^{-2A(z)} dz^2 + ds^2(M_6) ~,
\end{equation}
and the metric on $M_6$ is taken to be conformal to the round metric on the six-sphere $S^6$:
\begin{equation}
ds^2(M_6) = e^{2Q(z)} ds^2(S^6) ~.
\end{equation}
The non-zero torsion classes of the $SU(3)$-structure are
\begin{equation}
\R W_1 = 2 e^{-Q} ~, \qquad \R E = - \frac{9}{2} \left(\frac{q}{p}\right)^{-1/3}  e^A \frac{dQ}{dz} ~.
\end{equation}
It follows that $\hat J \equiv e^{-2Q} J$ and $\hat \O \equiv e^{-3Q} \Omega$ define a nearly-K\"{a}hler structure on $S^6$:
\begin{equation}
d \hat J = 3 \I \hat \O ~, \qquad d \hat \Omega = 2 \hat J \wedge \hat J ~.
\end{equation}
Setting $m=1$ as in \cite{Dibitetto:2018ftj} the solution is determined by
\begin{equation}
e^{2Q} = \left(\frac{q}{p}\right)^{1/3} \frac{1}{\sqrt{z}} ~,\qquad
e^{2A} = \frac{4}{9} \left(\frac{q}{p}\right)^{1/3} \frac{1+z^3}{\sqrt{z}} ~, \qquad
e^{\phi} = q^{-1/6} p^{-5/6} z^{-5/4} ~,
\end{equation}
and
\begin{equation}
\cos\theta = \frac{2}{3} e^{Q-A} ~.
\end{equation}
The only non-zero fluxes are $F_0$ and $F_6 = 5q\vol(S^6)$.

This solution arises as a near-horizon limit of a D2-O8 configuration. The internal space is non-compact, with $z \in [0, \infty]$. Near $z = 0$, there is an O8-plane singularity, and at $z \to \infty$ a type of D2-brane singularity; see \cite{Dibitetto:2018ftj} for more details. Flux quantization imposes $2 \pi F_0 \in \mathbb{Z}$, and $q/(6 \pi^2) \in \mathbb{Z}$.

\subsubsection*{AdS$_3 \times S^3 \times S^3 \times S^1$ with $\mathcal{N} = (4,0)$ supersymmetry}

The AdS$_3 \times S^3 \times S^3 \times S^1$ solution of \cite[sec. 4.1]{Macpherson:2018mif} belongs to the class of solutions with strict $SU(3)$-structure, i.e.\ $\theta = 0$. The warp factor $A$, and the dilaton $\phi$ satisfy
\begin{equation}
d_6 A = 0 ~, \quad d_6 \phi = 0 ~.
\end{equation}
The one-form $v$ is closed and locally is set to $v = e^{\phi_0} e^{3A(\rho)} d\rho$, where $\phi_0$ is a constant.
The metric on $M_7$ reads
\begin{equation}
ds^2(M_7) =  e^{2\phi_0} e^{6A(\rho)} d\rho^2 + e^{2A(\rho)} \widehat{ds^2}(M_6) ~.
\end{equation}
Furthermore, the torsion classes of the $SU(3)$-structure are restricted so that
\begin{subequations}
\begin{align}
d_6 \hat{J} &= \frac{3}{2} m \I \hat{\O} + \hat{W_3} ~, \\
d_6 \hat{\O} &= m \hat{J} \wedge \hat{J} ~,
\end{align}
\end{subequations}
where $\hat{J} \equiv e^{-2A} J$, $\hat \O \equiv e^{-3A} \O$, and $\hat{W}_3 \equiv e^{-2A} W_3$. The dilaton $\phi$ and the warp factor $A$ are given by:
\begin{equation}
e^\phi = e^{\phi_0} e^{5A} ~, \qquad
e^{-8A} = 2  F_0 e^{2\phi_0} \rho + \ell ~,
\end{equation}
where $\ell$ is a constant. The only non-zero fluxes are $F_0$ and
\begin{subequations}
\begin{align}
F_4 &= d\rho \wedge \left( 2 \I(\hat{W}_3^{(2,1)}) - \tfrac12 m  \R \hat\O \right) ~, \\
F_6 &= \frac{2}{3!} m e^{-\phi_0} \hat{J}^3 ~.
\end{align}
\end{subequations}
The Bianchi identity to be satisfied is that of $F_4$, $dF_4 = 0$, which yields
\eq{
d \I (\hat{W}_3^{(2,1)}) - \frac{m^2}{4} \hat{J} \wedge \hat{J} = 0 ~.
}
Thus, what remains to be done in order to solve the equations of motion is to find a suitable manifold $M_6$ admitting an $SU(3)$-structure with the right torsion classes.
On $S^3 \simeq SU(2)$, a set of left-invariant forms $(\sigma^{1}, \sigma^{2}, \sigma^{3})$ can be found such that
\eq{\label{liforms}
d \sigma^j &= \frac12 \e_{jkl} \sigma^k \wedge \sigma^l ~,
}
where $j, k, l \in \{ 1,2,3 \}$, and $\epsilon^{jkl}$ is the Levi--Civita symbol. Let $\sigma^j_a$ be the left-invariant forms on $S^3_a$, $a \in \{1,2\}$, and let
\eq{
\z^j &\equiv \frac{1}{\sqrt{2} m} (\sigma^j_2 + i \sigma^j_1 ) \\
\hat{J} &= \frac{i}{2} \delta_{ij} \z^i \wedge \bar{\z}^j ~, \qquad
\hat{\O} = \frac{1}{3! \sqrt{2}}(1+ i) \e_{jkl} \z^j \wedge \z^k \wedge \z^l ~.
}
Then $(\hat{J}, \hat{\O})$ form an $SU(3)$-structure on $S^3_1 \times S^3_2$, with corresponding metric
\eq{
\widehat{ds^2}(M_6) &= \frac{2}{m^2} \left( ds^2(S^3_1) + ds^2(S^3_2) \right) ~, \\
}
where $ds^2(S^3_a) = \frac{1}{4} \sum_j (\sigma_a^j)^2$. Making use of \eqref{liforms}, the non-vanishing torsion classes are determined to be
\eq{
\hat{W}_1 &= m ~,\qquad
\hat{W}_3^{(2,1)} = - m \frac{1+i}{4! \sqrt{2}} \e_{ijk} \z^i \wedge \z^j \wedge \bar{\z}^k ~,
}
with $\hat{W}_3^{(2,1)}$ satisfying
\eq{
d \hat{W}_3^{(2,1)} = \frac{m^2}{4} i \hat{J} \wedge \hat{J}~.
}
as desired. The fluxes now read:
\begin{subequations}
\begin{align}
F_4 &= \frac{4}{m^2} \left( \vol(S_1^3) + \vol(S_2^3) \right) \wedge d\rho ~, \\
F_6 &= \frac{16}{m^5} e^{-\phi_0} \vol(S_1^3) \wedge \vol(S_2^3) ~.
\end{align}
\end{subequations}
The coordinate $\rho$ here is related to the coordinate $r$ in \cite{Macpherson:2018mif} via
\begin{equation}
(2 F_0 e^{2\phi_0} \rho + \ell)^{1/2} = \frac{1}{L^{4}}(F_0 \nu r + c) ~,
\end{equation}
and also $e^{-\phi_0} = q L^5$, where $(L, \nu = \pm 1, c, q)$ are constant parameters in \cite{Macpherson:2018mif}.

\section{Type IIB}\label{IIB}

In this section we analyze the Type IIB supersymmetry equations \eqref{SUSY} (lower sign) in a way similar to that of the analysis of the Type IIA supersymmetry equations in the previous section.

The NSNS sector is constrained by \eqref{SUSYa}, which yields:
\begin{subequations}\label{IIBNSNS}
\begin{align}
d \left( e^{2A - \phi} \sin\t \right) &= 0 ~,   \label{warp_constraint} \\
d \left( e^{2A - \phi} \cos\t J \right) - e^{2A-\phi} \sin\t H &= 0 ~,   \\
d \left( e^{2A - \phi} \left(v \wedge \R \O - \sin\t \tfrac{1}{2} J^2\right)\right) - e^{2A-\phi} \cos\t H \wedge J &= 0 ~, \\
d \left(e^{2A - \phi} \cos\t J^3 \right) + 3! \, e^{2A- \phi}H \wedge \left(v \wedge \R \O - \sin\t \tfrac{1}{2} J^2 \right) &= 0 ~.
\end{align}
\end{subequations}
The RR field-strengths are derived from \eqref{SUSYb}, corresponding to
\begin{subequations}\label{IIBRR}
\begin{align}
e^{3A} \star_7 F_7 &= - 2 m e^{2A-\phi} \sin\theta ~, \\
e^{3A} \star_7 F_5 &= d\left(e^{3A-\phi}\cos\theta v\right) + 2 m e^{2A-\phi} \cos\theta J ~, \\
e^{3A} \star_7 F_3 &= d\left(e^{3A-\phi}\left(\sin\theta v \wedge J-\I\O\right)\right) + e^{3A-\phi}\cos\theta H \wedge v \nn \\
&- 2 m e^{2A-\phi}\left(v \wedge \R\O - \sin\theta \tfrac{1}{2} J^2\right) ~, \\
e^{3A} \star_7 F_1 &= -d\left(e^{3A-\phi}\cos\t v \wedge \tfrac{1}{2}J^2\right)
+ e^{3A-\phi} H \wedge \left(\sin\t v \wedge J-\I\O\right) \nn \\
&- m e^{2A-\phi} \cos\t \tfrac{1}{3} J^3 ~.
\end{align}
\end{subequations}

From \eqref{IIBNSNS}, in addition to $e^{2A-\phi} \sin\t$ being constant we obtain
\begin{subequations}\label{IIBH}
\begin{align}
\sin\t H &= \tfrac32 \cos\t \I ( \overline{W_1} \O) + \left[\cos\t d_6(2A - \phi) - \sin\t d_6 \t + \cos\t W_4 \right] \wedge J
+ \cos\t W_3 \nn \\
&+ v \wedge \left[\cos\t T_2 + \left(\tfrac23 \cos\t  \R E + \cos\t (2 \dot{A} - \dot{\phi}) - \sin\t \dot{\t}\right) J
+ \cos\t \R (\overline{V_2} \lrcorner \O) \right] ~,  \\
\cos\t H &= \cos\t H^R \R\O + \cos\t H^I \I\O + (2 \R V_1 -\sin\t W_4)\wedge J + 2 \cos\t \R(H^{(2,1)}) \nn \\
&+ v \wedge  \left[ -\R W_2 -\sin\t T_2 - \left(\tfrac{2}{3} \sin\theta \R E + \R W_1 \right) J \right] \nn \\
&+ v \wedge \I\left[(d_6(2A-\phi) - W_0 +  \overline{W_5}) \lrcorner \O \right] -  \sin\t v \wedge \R(\overline{V_2} \lrcorner \O) ~, \\
2H^I &= \cos\theta(2\dot A-\dot\phi)-\sin\theta \dot\t ~.
\end{align}
\end{subequations}
From \eqref{IIBRR}, making use of \eqref{IIBH} and the identities in the appendix of \cite{Passias:2019rga} to Hodge dualize, we derive the following expressions for the magnetic RR field-strengths:
\begin{subequations}\label{IIBR}
\begin{align}
e^\phi F_1 &= -\left(2\cos\t me^{-A}+4H^R+3\cos\t R\right) v + 2\I (X_1^{(1,0)}) ~,\\
e^\phi F_3 &= \left(-2me^{-A}-\cos\t H^R-\I E+\tfrac{3}{2}\sin\t\I W_1\right) \I \O \nn \\
&+ \left[- 2 \I V_1 -2\R V_2 + 2\sin\t\I(W_0^{(1,0)}-dA^{(1,0)})\right] \wedge J \nn \\
&+ v \wedge (\I W_2-\sin\t T_1)
-2\left[\I W_1 - \sin\t(R+me^{-A})\right] v \wedge J\nn \\
&+ 2 \cos\t \I(H^{(2,1)}) + 2 \sin\t \I(W_3^{(2,1)}) -  \R S + X_3 \lrcorner(v \wedge \R \O)~, \\
e^\phi F_5 &=  \cos\t \left( R + 2m e^{-A}\right) v  \wedge \tfrac12 J^2 - \I ( X_5^{(1,0)}) \wedge J^2   \nn \\
&- \cos\t v \wedge J \wedge T_1 + 2 \cos\t v \wedge  \R V_1  \wedge \I \O ~, \\
e^{\phi} F_7 &= - 2me^{-A}\sin\theta \vol_7 ~,
\end{align}
\end{subequations}
where
\begin{subequations}
\begin{align}
X_1 &\equiv -\cos\t d_6(A-\phi) + \sin\t d_6\t - \cos\t W_0 - 8 \I(H_v^{(1,0)}) ~,\\
X_3 &\equiv dA + d_6(2A-\phi) + W_5 + \overline{W_5} - 2 \sin\theta \R V_1 ~, \\
X_5 &\equiv \cos\t d_6(3A-\phi) - \cos\t W_0 - \sin\t d_6\theta ~.
\end{align}
\end{subequations}
Substituting the above expressions in the pairing equation \eqref{SUSYc} yields the scalar constraint
\begin{equation}\label{IIBpairing}
3R + 6me^{-A} + 4\cos\t H^R + 2\I E - 6 \sin\t \I W_1 = 0 ~.
\end{equation}

Equations \eqref{warp_constraint}, \eqref{IIBH}, \eqref{IIBR}, and \eqref{IIBpairing} constitute necessary and sufficient conditions for the
preservation of supersymmetry.

\subsection{Solutions}\label{IIBsol}

A family of solutions for the limiting case $\t = 0$, that is the strict $SU(3)$-structure case, were examined in \cite[sec. 5]{Passias:2019rga}: the internal manifold $M_7$ is a $U(1)$ fibration over a conformally K\"{a}hler base, and they feature a varying axio-dilaton, a primitive $(2,1)$-form flux $H + i e^\phi F_3$, and five-form flux $F_5$.  The solutions of \cite{Kim:2005ez, Donos:2008ug, Benini:2013cda, Benini:2015bwz, Couzens:2017nnr}, with $\mathcal{N} = (2,0)$ supersymmetry,  belong in this family.

Here, we will examine the other limiting case: $G_2$-structure solutions, i.e., solutions with $\t = \pi/2$. Although equivalent, it turns out to be more convenient to work directly with the $G_2$-structure rather than to use the $\theta = \pi/2$ limit of the supersymmetry conditions derived above.

The polyforms $\psi_\pm$ are parameterized in terms of the $G_2$-structure, defined by the three-form $\varphi$, as
\eq{
\psi_+ = \frac18 e^A \left( 1- \star_7 \vf\right) ~, \qquad \psi_- = \frac18 e^A \left(- \vf + \vol_7 \right) ~.
}
Plugging these expressions into the supersymmetry equations \eqref{SUSY}, and making use of \eqref{g2torsion} leads to the following constraints for the torsion classes
\eq{
\tau_1 = \tau_2 &= 0 ~,  \qquad \tau_0 = - \frac{12}{7}m e^{-A} ~.
}
Vanishing of the $\tau_2$ torsion class means that the $G_2$-structure is integrable, meaning one can introduce a $G_2$ Dolbeault cohomology \cite{Fernandez1998DolbeaultCF}.
Furthermore, we obtain
\eq{
d(2A-\phi) = 0 ~, \qquad H = 0 ~,
}
and
\eq{
e^\phi F_3 &= d A \lrcorner \star_7 \vf + \frac{2}{7} m e^{-A} \vf + \tau_3 ~, \\
e^\phi F_7 &= -2 m e^A \vol_7 ~,
}
while $F_1 = F_5 = 0$.

Next, we examine the Bianchi identities, which reduce to $d F_3 = 0$. Imposing the Bianchi identities in addition to the supersymmetry conditions yields a solution to the equations of motion.
We will work with a rescaled $G_2$-structure $\hat{\vf} = e^{-3A} \vf$ and corresponding metric $\widehat{ds^2}(M_7) = e^{-2A} ds^2(M_7)$.
The rescaled torsion classes are given by
\eq{
\hat{\tau}_0 = e^A \tau_0 = - \frac{12}{7} m ~, \qquad \hat{\tau}_1 = \tau_1 - dA = - dA ~, \qquad \hat{\tau}_2 = e^{-A} \tau_2 = 0 ~, \qquad \hat{\tau}_3 = e^{-2A} \tau_3~.
}
Using these, the Bianchi identities read
\eq{\label{eom}
d \left( \hat{\tau}_3 - \frac16 \hat{\tau}_0 \hat{\vf} -\hat{\tau}_1 \lrcorner \hat{\star}_7 \hat{\vf} \right) = 0 ~.
}
Thus, the problem of finding a solution to the equations of motion is reduced to this purely geometric condition. Note that (up to constant prefactors), the same condition appears for heterotic backgrounds on $G_2$-structure spaces \cite[eq. (2.13)]{delaOssa:2014lma}.

\subsubsection{Examples}
Let us now give several examples of solutions to the Bianchi identities, which have been reduced to the constraint \eqref{eom}.

First, we consider $M_7 = S^3 \times M_4$, with standard $G_2$-structure, trivial warp factor $A=0$,  and where $M_4$ is any hyper-K\"{a}hler manifold \cite{Prins:2018hjc}.
This recovers the near-horizon limit of D1- and D5-branes, with $\mathcal{N}=(4,4)$ supersymmetry \cite{Maldacena:1997re}.
Let $(\s^{1},\s^{2},\s^{3})$ be the left-invariant one-forms on $S^3$ satisfying $ d \s^j = \frac12 \e^{jkl} \s^k \wedge \s^l$ and $(\o^{1},\o^{2},\o^{3})$ be the hyper-K\"{a}hler structure on $M_4$ satisfying
\begin{equation}
d \o^j = 0 ~, \qquad \frac{1}{2} \omega_i \wedge \omega_j = \delta_{ij}\vol(M_4) ~.
\end{equation}
Then the $G_2$-structure
\eq{
8 m^3 \hat{\vf} &= - \vol(S^3) -  \frac{1}{2\sqrt{2}} \delta_{ij }\o^i \wedge \s^j \\
16 m^4 \hat{\star}_7 \hat{\vf} &=  \vol(M_4) + \frac{1}{8\sqrt{2}} \epsilon_{jkl} \o^j \wedge \s^k \wedge \s^l
}
has
\eq{
\hat{\tau}_0 &= -\frac{12}{7} m  ~, \qquad  4 m^2 \hat{\tau}_3 = - \frac67  \vol(S^3) + \frac{1}{14\sqrt{2}}\o^j \wedge \s^j ~,
}
and $\hat{\tau}_1 = \hat{\tau}_2$ = 0, which satisfy \eqref{eom}.

The next two examples are solutions in the presence of spacetime-filling O5-plane and D5-brane sources, which wrap calibrated three-cycles inside $M_7$. The presence of these lead to a source term in the Bianchi identity, $d F_3 = J_4$. This thus modifies the right-hand side of \eqref{eom} such that the sourced Bianchi identities instead reduce to
\eq{\label{eom_sources}
d \left( \hat{\tau}_3 - \frac16 \hat{\tau}_0 \hat{\vf} -\hat{\tau}_1 \lrcorner \hat{\star}_7 \hat{\vf} \right) = J_4 ~.
}

The first sourced example is given by the twisted toroidal orbifold $M_7 = T^7 / (\zbb_2 \times \zbb_2 \times \zbb_2)$:
we refer the reader to \cite{DallAgata:2005zlf} which we follow closely, as well as \cite{Emelin:2021gzx} for details.
Given a set of coordinates $y^m$ on $M_7$, we may introduce a twisted frame $\{e^m(y)\}$. In terms of this frame, the three-form determining the $G_2$-structure can then be defined as
\eq{\label{explicit_g2}
\vf &= e^{127} - e^{347} - e^{567} + e^{136} - e^{235} + e^{145} + e^{246} \\
\star \vf &= e^{3456} - e^{1256} - e^{1234} + e^{2457}- e^{1467} + e^{2367} + e^{1357}~.
}
Generically, the frame satisfies
\eq{
d e^m = \frac12 \tau^m_{np} e^n \wedge e^p~.
}
This twisting breaks the $G_2$-holonomy of the toroidal orbifold by introducing non-vanishing torsion classes $\tau_0$, $\tau_3$, such that $\vf$ is co-closed, but no longer closed. The representation of the $\zbb_2$-involutions on the coordinates $y^m$ of $M_7$, as well as the consistency constraint $d^2 e^m = 0$, restrict the possible values $\tau_{np}^m$ can take. We will restrict our attention to $\tau_{np}^m$ being the structure constants of $SO(p,q) \times U(1)$ with $p+q = 4$: this comes down to setting
\eq{
\left(\begin{array}{llll}
 \tau^{1}_{45} & \phantom{-}\tau^{2}_{46}   & \phantom{-}\tau^{1}_{36}  & - \tau^{2}_{35}  \\
 \tau^{3}_{25} & -  \tau^{3}_{16}           & - \tau^{4}_{26}           & - \tau^{4}_{15}  \\
 \tau^{6}_{24} & \phantom{-}\tau^{5}_{14}   & - \tau^{5}_{23}           &  \phantom{-} \tau^{6}_{13}
\end{array}
\right)
=
\left(\begin{array}{llll}
\phantom{-}a_4 \frac{a_1}{a_2}              & \phantom{-}a_5 \frac{a_1}{a_3}            & - a_6 \frac{a_1}{a_2}                     & a_1 \\
\phantom{-}a_4 \phantom{\frac{a_1}{a_2} }   & - a_5 \frac{a_2}{a_3}                     & \phantom{-}a_6 \phantom{\frac{a_1}{a_2}}  & a_2 \\
- a_4 \frac{a_3}{a_2}                       & \phantom{-}a_5 \phantom{\frac{a_1}{a_2}}  & - a_6 \frac{a_3}{a_2}                     & a_3
\end{array}
\right)~,
}
with $a_i$ constant and all other $\tau^m_{np}$ vanishing.
Neither $\tau_0$ nor $\tau_3$ vanishes generically, with
\eq{
\tau_0 = - \frac27 \left(-a_1 - a_2 + a_3 + a_4 + a_5 - a_6
+ \frac{a_1 a_4}{a_2} - \frac{a_1 a_6}{a_2} - \frac{a_3 a_6}{a_2}
+ \frac{a_1 a_5}{a_3} - \frac{a_2 a_4}{a_3} + \frac{a_2 a_5}{a_3}\right)~.
}
As discussed in \cite{Emelin:2021gzx}, setting $A=0$ leads to solutions with source term $J_4$ given by
\eq{
J_4 = k_m (a_i) \psi^m ~,
}
with $\psi^m \in \{e^{3456}, e^{1256}, e^{1234}, e^{2457}, e^{1467}, e^{2367}, e^{1357} \}$ and $k_m(a_i)$ dependent on the twisting parameters. Note that the AdS$_3$ radius is proportional to the torsion class $\tau_0(a_i)$, and hence the twisting parameters $a_i$ are restricted such that $\tau_0 \neq 0$.

The second example of a sourced solution can be obtained by taking $M_7= H(3,1)$, the generalized Heisenberg group, as discussed in \cite{Fernandez:2008wla} and recently investigated in the context of three-dimensional heterotic Minkowski backgrounds \cite{delaOssa:2021cgd}.
Geometrically, $H(3,1)$ is a nilmanifold, for which a frame can be found satisfying
\eq{
d e^m = \left\{ \begin{array}{ll}
0 \qquad                        &m \neq 7 \\
a e^{12} + b e^{34} + c e^{56} & m=7
\end{array}\right. ~,
}
with $a,b,c$ non-zero parameters. Again expressing the three-form $\varphi$ in terms of the frame $\{ e^m \}$ as in \eqref{explicit_g2},
it follows that $\tau_1 = \tau_2 = 0$ and
\eq{
\tau_0 = \frac27 (a+b+c)~,
}
which restricts the parameters to satisfy $a+ b+ c \neq 0$ in order to find AdS$_3$ backgrounds.\footnote{Compared to \eqref{explicit_g2}, we reparametrize $e^4 \rightarrow - e^4$, $e^5 \leftrightarrow e^6$ to remain consistent with \cite{Fernandez:2008wla}.} Setting $A=0$, one finds \eqref{eom_sources} is satisfied with calibrated source term $J_4$ given by
\eq{
J_4 &= k_1 e^{1234} + k_2 e^{3456} + k_3 e^{1256}
}
with $k_1 = (a+b)^2 + (ab + bc + ca)$ and $(a,b,c)$ cyclically permuted for $k_2, k_3$.

\vskip 1cm

\noindent
{\bf Acknowledgements}

\noindent
We would like to thank G.~Lo  Monaco, N.~Macpherson, E.~Svanes and A.~Tomasiello for useful discussions. The work of A.P. is supported by the LabEx ENS-ICFP: ANR-10-LABX-0010/ANR-10-IDEX-0001-02 PSL*. The work of D.P. is supported by the Groenvelder Institute for Applied Hydrodynamics and Turbulence (GIAHT).

\appendix

\bibliography{3}
\bibliographystyle{utphys}

\end{document}